\documentclass[preprint,aps,floats,prb,fancyhdr,a4paper,superscriptaddress,nofootinbib,showpacs]{revtex4}
\usepackage{latexsym}
\usepackage{amsmath}
\usepackage{amssymb}
\usepackage{amsfonts}
\usepackage{graphicx}
\usepackage{dcolumn}
\usepackage{fancyhdr}
\usepackage[colorlinks]{hyperref}

\begin{document}

\title{PERFECT FLUID COSMOLOGICAL MODELS WITH TIME-VARYING CONSTANTS}

\author{\footnotesize JOS\'E ANTONIO BELINCH\'{O}N\footnote{e-mail: abelcal@ciccp.es}}

\address{Grupo Inter-Universitario de An\'{a}lisis Dimensional. \\
Dept. F\'{\i}sica ETS Arquitectura UPM Av. Juan de Herrera 4. Madrid
28040 Espa\~{n}a.}

\author{\footnotesize INDRAJIT CHAKRABARTY}

\address{Blk 278, \#06-183, Toh Guan Road, Singapore 600278}

\begin{abstract}
In this paper, we study in detail a perfect fluid cosmological model with 
time-varying ``constants" using dimensional analysis and the symmetry method. 
We examine the case of variable ``constants" in detail without considering 
the perfect fluid model as a limiting case of a model with a causal bulk 
viscous fluid as discussed in a recent paper. We obtain some new solutions 
through the Lie method and show that when matter creation is considered,
these solutions are physically relevant. 
\end{abstract}

\maketitle

\section{Introduction}

Cosmological models with time-varying ``constants" have been studied for quite 
some time ever since Dirac$^{1}$ proposed a theory with a time-varying 
gravitational constant $G$. Several works have investigated cosmological models
with variable cosmological constant within a framework of dissipative 
thermodynamics$^{2-14}$. In a recent paper$^{15}$, we have studied a 
causal bulk viscous cosmological model with time-varying constants. We arrived
to the conclusion that our cosmological model tends to a perfect fluid one 
in the matter predominance era. The approach of our work was to study the 
symmetries of the model beginning with the dimensional analysis of the 
field equations. The method of Lie group allowed us to arrive to the 
conclusion that under the hypotheses considered there is only one solution 
for our model, the scaling one that was trivially obtained using dimensional 
analysis.

Since our viscous model tends to a perfect fluid one, the purpose of this work
is to perform a detailed study of all the possible symmetries of a perfect
fluid model with time varying constants showing that in this case it is
possible to find more solutions in addition to the scaling one. In order 
to carry out this study, we begin in section 2 by outlining the equations 
that govern the model as well as the notation employed. In section 3, 
we review the scaling solution obtained in previous works highlighting
the ``assumed'' hypotheses that we need to make in order to obtain a 
solution using dimensional analysis, these are: $div(T_{j}^{i})=0,$ 
conservation principle, and that the relation $G/c^{2}$ remain constant 
for all value of $t$ (cosmic time). We emphasize the special case $\omega=0$ 
i.e. the dust case as this is the scenario that describes our model in the
matter predominance era (in agreement with our previous paper). We discuss 
some interesting relationships that arise with the similarity method.

In Section 4, we work towards finding other possible solutions to the field 
equations using the Lie group method. We start this section by rewriting 
the field equations in such a way that we can use the standard Lie procedure 
that allow us to find more symmetries. After outlining the equation and
the constraint, we proceed to study some cases. The first one is the obtained
previously using dimensional analysis since dimensional analysis is just a special
class of symmetry. We would like to emphasize that the Lie method show us that
one of the assumptions made with the dimensional method, $G/c^{2}=const.,$ is
at least correct from the mathematical point of view. This result allow us to
validate completely the solution obtained through similarity. This solution 
connects perfectly with our previous work$^{15}$ 

In order to show that the other two solutions are physically relevant, we 
examine them in a matter creation scenario in section 5. We find that 
the horizon and entropy problems are solved when we consider matter creation
and this leads to a set of physically relevant solutions. We note that 
no new constants or assumptions are required for this exercise. We conclude the 
paper by summarizing the results in section 6.

\section{The Model\label{Model}}

We will use the field equations in the form:
\begin{equation}
\label{ECU1}
R_{ij}-\frac{1}{2}g_{ij}R=\frac{8\pi G(t)}{c^{4}(t)}T_{ij}+\Lambda(t)g_{ij},
\end{equation}
where the energy momentum tensor is:
\begin{equation}
T_{ij}=\left(  \rho+p\right)  u_{i}u_{j}-pg_{ij},%
\end{equation}
and $p=\omega\rho$ in such a way that $\omega\in\left[  0,1\right]  .$

The cosmological equations are now (for a flat FRW Universe as the most recent observations suggest us$^{16-19}$.):
\begin{eqnarray}
2H^{\prime}+3H^{2}  & = &-\frac{8\pi G(t)}{c(t)^{2}}p+c(t)^{2}\Lambda
(t),\label{p1}\\
3H^{2}  &  = & \frac{8\pi G(t)}{\,c(t)^{2}}\rho+c(t)^{2}\Lambda(t), \label{p2}%
\end{eqnarray}
where $H=(f^{\prime}/f)$ is the Hubble function. Applying the
covariance divergence to the second member of equation (\ref{ECU1}) we get:
\begin{equation}
T_{i;j}^{j}=\left(  \dfrac{4c_{,j}}{c}-\dfrac{G_{,j}}{G}\right)  T_{i}%
^{j}-\dfrac{c^{4}(t)\delta_{i}^{j}\Lambda_{,j}}{8\pi G},%
\end{equation}
which simplifies to:
\begin{equation}
\rho^{\prime}+3(\omega+1)\rho H=-\dfrac{\Lambda^{\prime}c^{4}}{8\pi G}%
-\rho\dfrac{G^{\prime}}{G}+4\rho\dfrac{c^{\prime}}{c}.%
\end{equation}
We assume that $div(T_{j}^{i})=0$ which leads to the two following
equations:
\begin{eqnarray}
\rho^{\prime}+3(\omega+1)\rho H  & = &0,\\
-\dfrac{\Lambda^{\prime}c^{4}}{8\pi G}-\rho\dfrac{G^{\prime}}{G}+4\rho
\dfrac{c^{\prime}}{c}  & = &0.
\end{eqnarray}

Hence the field equations are:
\begin{eqnarray}
2H^{\prime}+3H^{2}  & = & -\frac{8\pi G}{c^{2}}p+\Lambda c^{2},\label{field1}\\
3H^{2}  & = & \frac{8\pi G}{c^{2}}\rho+\Lambda c^{2},\label{field2}\\
\rho^{\prime}+3\left(  \omega+1\right)  \rho H  & = & 0,\label{field3}\\
-\frac{\Lambda^{\prime}c^{4}}{8\pi\rho G}-\frac{G^{\prime}}{G}+4\frac
{c^{\prime}}{c}  & = & 0. \label{field4}%
\end{eqnarray}

\section{Review of the scaling solution}

As have been pointed out by Carr and Coley$^{20}$, the existence of self-
similar solutions (Barenblatt and Zeldovich$^{21}$) is related to
conservation laws and to the invariance of the problem with respect to the
group of similarity transformations of quantities with independent dimensions.
This can be characterized within general relativity by the existence of a
homotetic vector field and for this reason one must distinguish between
geometrical and physical self-similarity. Geometrical similarity is a property
of the spacetime metric, whereas physical similarity is a property of the matter
fields (our case). In the case of perfect fluid solutions admitting a
homotetic vector, geometrical self-similarity implies physical self-similarity.

As we show in this section as well as in previous works, the assumption of
self-similarity reduces the mathematical complexity of the governing
differential equations. This makes such solutions easier to study
mathematically. Indeed self-similarity in the broadest Lie sense refers to an
invariance which allows such a reduction.

Perfect fluid space-times admitting a homotetic vector within general
relativity have been studied by Eardley$^{22}$. In such space-times, all
physical transformations occur according to their respective dimensions, 
in such a way that geometric and physical self-similarity coincide. 
It is said that these space-times admit a transitive similarity group 
and space-times admitting a non-trivial similarity group are called 
self-similar. Our model i.e. a flat FRW model with a perfect fluid 
stress-energy tensor has this property and as already have been pointed 
out by Wainwright$^{23}$, this model has a power law solution.

Under the action of a similarity group, each physical quantity $\phi$
transforms according to its dimension $q$ under the scale transformation. For
space-times with a transitive similarity group, dimensionless quantities are
therefore spacetime constants. This implies that the ratio of the pressure of
the energy density is constant so that the only possible equation of state is
the usual one in cosmology i.e. $p = \omega\rho$, where $\omega$ is a constant.
In the same way, the existence of homotetic vector implies the existence of
conserved quantities.

In this section we would like to review the solution obtained through
Dimensional Analysis$^{24-25}$. Our starting point is the 
condition $div\left(  T_{i}^{j}\right) = 0$ which allows us to obtain one of
the dimensional constants that we need in order to apply the method of 
dimensional analysis. Therefore, from eq. (\ref{field3})
\begin{equation}
\rho^{\prime}+3\left(  \rho+p\right)  H=0,
\end{equation}
we obtain the following relation between the energy density and the scale
factor as well as the constant of integration that we shall need for our 
subsequent calculations:
\begin{equation}
\rho=A_{\omega}f^{-3(\omega+1)}, \label{e4}%
\end{equation}
where $A_{\omega}$ is the integration constant that depends on the equation of
state that we want to consider i.e. constant $\omega$, $\left[  A_{\omega
}\right]  = L^{3(\omega+1)-1}MT^{-2}$. We consider a second dimensional constant
by considering the relation $G/c^{2}=B,$ where, $B$ is the constant. This is
a hypothesis which is necessary in order to apply dimensional analysis. In this
next section, we will see that this condition on $G$ and $c$ is mathematically
correct. Our purpose here is to show that no more hypotheses are necessary to 
solve the differential equations that govern the model (see $^{26-28}$ for 
standard text-book on Dimensional Analysis and $^{29}$ for applications to Cosmology).

Therefore, if we take into account the standard dimensional procedure, we find
that the set of governing parameters are $M=M\left\{  A_{\omega},B,t\right\}
,$ which bring us to obtain the next relations:
\begin{equation}%
\begin{array}
[c]{l}%
G\propto A_{\omega}^{\frac{2}{\gamma+1}}B^{\frac{2}{\gamma+1}+1}%
t^{\frac{2(1-\gamma)}{\gamma+1}},\\
c\propto A_{\omega}^{\frac{1}{\gamma+1}}B^{\frac{1}{\gamma+1}}t^{\frac
{(1-\gamma)}{\gamma+1}},\\
\rho\propto B^{-1}t^{-2},\\
f\propto A_{\omega}^{\frac{1}{\gamma+1}}B^{\frac{1}{\gamma+1}}t^{\frac
{2}{\gamma+1}},\\
k_{B}\theta\propto A_{\omega}^{\frac{3}{\gamma+1}}B^{\frac{3}{\gamma+1}%
-1}t^{\frac{4-2\gamma}{\gamma+1}},\\
\Lambda\propto A_{\omega}^{\frac{-2}{\gamma+1}}B^{\frac{-2}{\gamma+1}}%
t^{\frac{-4}{\gamma+1}},\\
q=\frac{\gamma-1}{2},%
\end{array}
\label{Tabla1}%
\end{equation}
where $\gamma=3(\omega+1)-1,$ and $q$ is the deceleration parameter.

From the set of equations (\ref{Tabla1}), we see that $\frac{G}{c^{2}%
}=B$ (trivially), $f=ct$ (the horizon problem is missing in this model),
$\Lambda\propto\frac{1}{c^{2}t^{2}}\propto f^{-2}$ for all value of $\gamma$
i.e. $\forall\omega$.

We would like to point out that the same results have been obtained by Midy
and Pettit in a more general context$^{30}$.

As we have indicated in the introduction, in a recent paper$^{15}$ we have
studied a full causal bulk viscous cosmological model with time varying
constants. The main conclusions in that paper was that under the ``assumed''
hypotheses, i.e. $div(T_{j}^{i})=0,$ conservation principle, and $\Pi
\propto\rho,$ the bulk viscous pressure behaves as the energy density and we
obtained the following results:

\begin{enumerate}
\item  there is only one solution for the field equations (scaling solution),

\item  the bulk viscous pressure behaves as an adiabatic matter mechanism
(solving the entropy problem),

\item  the ``constants'' $G,c$ and $\Lambda$ are decreasing functions on time
(solving the horizon problem) and

\item  in the matter predominance era, i.e. $\omega=0,$ we cannot consider the
bulk viscosity and therefore our viscous fluid tends to a perfect fluid one
(see $^{15}$ for details).
\end{enumerate}

Due to these conclusions, in the present paper, we have studied a perfect fluid
model with variable ``constant" emphasizing the case of matter predominance
era or a ``dust" solution; in other words, considering $\omega=0$ in the equation
of state. In this case, we obtain the following solutions:%
\begin{eqnarray}
G  & \propto & t^{-2/3}, c\propto t^{-1/3}, \rho\propto
t^{-2}\nonumber\\
f  & \propto & t^{2/3}, \theta\propto t^{0}, \Lambda\propto
t^{-4/3},q = 1/2 \label{PF}%
\end{eqnarray}
while in the viscous model, the solutions are:%
\begin{eqnarray}
G  & \propto & A_{\omega,\varkappa}^{\frac{2}{\alpha+1}}k_{\gamma}%
^{\frac{3+\alpha}{b\left(  \alpha+1\right)  }}t^{-4-\frac{3+\alpha}{b\left(
\alpha+1\right)  }},\label{g}\\
c  & \propto & A_{\omega,\varkappa}^{\frac{1}{\alpha+1}}k_{\gamma}^{\frac
{1}{b\left(  \alpha+1\right)  }}t^{-1-\frac{1}{b\left(  \alpha+1\right)  }%
},\label{c}\\
\rho & \propto & k_{\gamma}^{-b^{-1}}t^{b^{-1}}\propto\Pi,\label{spi}\\
f  &  \propto & A_{\omega,\varkappa}^{\frac{1}{\alpha+1}}k_{\gamma}^{\frac
{1}{b\left(  \alpha+1\right)  }}t^{-\frac{1}{b\left(  \alpha+1\right)  }%
},\label{entro}\\
\Lambda & \propto & A_{\omega,\varkappa}^{\frac{-2}{\alpha+1}}k_{\gamma}%
^{\frac{-2}{b\left(  \alpha+1\right)  }}t^{\frac{2}{b\left(  \alpha+1\right)
}}, \label{q}%
\end{eqnarray}
where, $\alpha=3(\omega+1+\varkappa)-1$, $b=\gamma-1,$(see $^{15}$ for
details). If we consider $\gamma=1/2,$ $\omega=0$ and $\varkappa=0$ (viscous
pressure vanishes and therefore $k_{\gamma}$ vanishes too) then
we obtain the same results as in equation (\ref{PF}), but, as we have indicated
previously this approach is inconsistent when $\omega=0$ and for this reason we
have studied the perfect fluid case.

As we can see in the perfect fluid case, ``constants'' $G, c$ and
$\Lambda$ are decreasing functions on time, but in this case decrease slower
than in the radiation predominance era, while $\rho$ and $f$ behave as in the
FRW model.

To obtain these solutions we needed two dimensional constants, viz., 
$A_{\omega}$ and $B$. In the next section, we shall show that constant $B$ is a
reasonable hypothesis since with the Lie group method such condition holds as
a result in the scaling solution.

\section{Lie method}

As we have seen earlier, the $\pi-monomia$ is the main object in dimensional 
analysis. It may be defined as a product of quantities which are invariant 
under changes of fundamental units. $\pi-monomia$ are dimensionless quantities, 
their dimensions are equal to unity. Dimensional analysis has the structure 
of a Lie group$^{31}$. The $\pi-monomia$ are invariant under the action of the
similarity group. On the other hand, we must mention that the similarity group
is only a special class of the mother group of all symmetries that can be
obtained using the Lie method. For this reason, when one uses dimensional
analysis, only one of the possible solutions to the problem is obtained.

As we have been able to find a solution through dimensional analysis, it is
possible that there are other symmetries of the model, since dimensional
analysis is a reminiscent of scaling symmetries, which obviously are not the
most general form of symmetries. Hence, we shall study the model through the
method of Lie group symmetries, showing that under the assumed hypotheses
there are other solutions of the field equations. In this section we shall
show how the lie method allows us to obtain different solutions for the field
equations. In particular we seek the forms of $G$ and $c$ for which our field
equations admit symmetries i.e. are integrable (see $^{32-37}$).

An alternative use of the Lie groups have been performed by M. Szydlowki et. al.
$^{38-39}$ where they study the Friedman equations in order to find the 
correct equation of state following pionerr works of Collins$^{40}$. 

In order to use the Lie method, we rewrite the field equations as follows.
From (\ref{field1}) $-$ (\ref{field2}), we obtain
\begin{equation}
2\frac{f^{\prime\prime}}{f}-2\left(  \frac{f^{\prime}}{f}\right)  ^{2}%
=-\frac{8\pi G}{c^{2}}\left(  p+\rho\right)  ,
\end{equation}
and therefore
\begin{equation}
2\left(  H\right)  ^{\prime}=-\frac{8\pi G}{c^{2}}\left(  p+\rho\right)  .
\end{equation}
From equation (\ref{field3}), we can obtain
\begin{equation}
H=-\frac{\rho^{\prime}}{3\left(  \left(  \omega+1\right)  \rho\right)  },%
\end{equation}
therefore
\begin{equation}
\left(  \frac{\rho^{\prime}}{\rho}\right)  ^{\prime}=12\pi\left(
\omega+1\right)  ^{2}\frac{G}{c^{2}}\rho.
\end{equation}
Taking $12\pi\left(  \omega+1\right)  ^{2}=A$ and then expanding, we obtain
\begin{equation}
\rho^{\prime\prime}=\frac{\rho^{\prime2}}{\rho}+A\frac{G}{c^{2}}\rho^{2}.
\label{ber1}%
\end{equation}

Now, we apply the standard Lie procedure to this equation. A vector field
\ $X$
\begin{equation}
X=\xi(t,\rho)\partial_{t}+\eta(t,\rho)\partial_{\rho},%
\end{equation}
is a symmetry of (\ref{ber1}) iff
\[
-\xi f_{t}-\eta f_{\rho}+\eta_{tt}+\left(  2\eta_{t\rho}-\xi_{tt}\right)
\rho^{\prime}+\left(  \eta_{\rho\rho}-2\xi_{t\rho}\right)  \rho^{\prime2}%
-\xi_{\rho\rho}\rho^{\prime3}+...
\]%
\begin{equation}
...+\left(  \eta_{\rho}-2\xi_{t}-3\rho^{\prime}\xi_{\rho}\right)  f-\left[
\eta_{t}+\left(  \eta_{\rho}-\xi_{t}\right)  \rho^{\prime}-\rho^{\prime2}%
\xi_{\rho}\right]  f_{\rho^{\prime}}=0. \label{ber2}%
\end{equation}
By expanding and separating (\ref{ber2}) with respect to powers of
$\rho^{\prime}$, we obtain the overdetermined system:
\begin{eqnarray}
\xi_{\rho\rho}+\rho^{-1}\xi_{\rho}  & = & 0,\label{EDP1}\\
\eta_{\rho\rho}-2\xi_{t\rho}+\rho^{-2}\eta-\rho^{-1}\eta_{\rho}  &
= & 0,\label{EDP2}\\
2\eta_{t\rho}-\xi_{tt}-3A\frac{G}{c^{2}}\rho^{2}\xi_{\rho}-2\rho^{-1}\eta_{t}
& = & 0,\label{EDP3}\\
\eta_{tt}-A\left(  \frac{G^{\prime}}{c^{2}}-2G\frac{c^{\prime}}{c^{3}}\right)
\rho^{2}\xi-2\eta A\frac{G}{c^{2}}\rho+\left(  \eta_{\rho}-2\xi_{t}\right)
A\frac{G}{c^{2}}\rho^{2}  & = & 0. \label{EDP4}%
\end{eqnarray}

Solving (\ref{EDP1}-\ref{EDP4}), we find that
\begin{equation}
\xi(t,\rho)=-2et+a,\eta(t,\rho)=\left(  bt+d\right)
\rho,\label{ber3}%
\end{equation}
subject to the constrain
\begin{equation}
\frac{G^{\prime}}{G}=2\frac{c^{\prime}}{c}+\frac{bt+d-4e}{2et-a}, \label{ber4}%
\end{equation}
with $a,b,e,$ and $d$ as constants. In order to solve (\ref{ber4}), we consider 
the following cases.

\subsection{\textbf{Case I:} $b=0$ and $d-4e=0$}\label{sol1}

In this case, the solution (\ref{ber4}) reduces to 
\begin{equation}
\frac{G^{\prime}}{G}=2\frac{c^{\prime}}{c}\Longrightarrow\frac{G}{c^{2}%
}=B=const.
\end{equation}
which means that ``constants'' $G$ and $c$ vary but in such a way that the
relation $\frac{G}{c^{2}}$ remains constant.

The solution obtained through Dimensional Analysis needs to make this
relations as hypothesis in order to obtain a complete solution for the field
equations. This case shows us that such hypothesis is correct (at least has
mathematical sense).

The knowledge of one symmetry $X$ might suggest the form of a particular
solution as an invariant of the operator $X$ i.e. the solution of
\begin{equation}
\frac{dt}{\xi\left(  t,\rho\right)  }=\frac{d\rho}{\eta\left(  t,\rho\right)
}, \label{rho}%
\end{equation}
this particular solution is known as an invariant solution (generalization of
similarity solution), therefore the energy density is obtained as
\begin{equation}
\frac{dt}{-2et+a}=\frac{d\rho}{4e\rho}\Longrightarrow\rho=\frac{1}{\left(
2et-a\right)  ^{2}},%
\end{equation}
for simplicity we adopt
\begin{equation}
\rho=\rho_{0}t^{-2},%
\end{equation}

Once we have obtained $\rho$, we can obtain $f$ (the scale factor) from
\begin{equation}
\rho=A_{\omega}f^{-3\left(  \omega+1\right)  }\Longrightarrow f=\left(
A_{\omega}t\right)  ^{\frac{2}{3\left(  \omega+1\right)  }}, \label{f}%
\end{equation}
in this way we find $H$ and from eq. (\ref{field2}), we obtain the
behaviour of $\Lambda$ as:
\begin{equation}
c^{2}\Lambda=3H^{2}-\frac{8\pi G}{c^{2}}\rho,\label{lamda}%
\end{equation}
and therefore,
\begin{equation}
\Lambda=\left(  3\beta^{2}-8\pi B\rho_{0}\right)  \frac{1}{c^{2}t^{2}}%
=\frac{l}{c^{2}t^{2}}.%
\end{equation}
If we replace all these results into eq. (\ref{field4}), then we obtain
the exact behaviour for $c,$ i.e.,
\begin{equation}
-\left(  \frac{1}{t}+\frac{c^{\prime}}{c}\right)  \lambda=\frac{c^{\prime}}{c},%
\end{equation}
where $\lambda=\frac{l}{8\pi B\rho_{0}}$, with $\lambda\in\mathbb{R}^{+}$, i.e. is a positive real number  and thus,
\begin{equation}
c=c_{0}t^{-\alpha},
\end{equation}
with $\alpha=\left(  \frac{\lambda}{1+\lambda}\right)  .$

\begin{figure}[h]
\begin{center}
\includegraphics[height=1.612in,width=5.3454in]{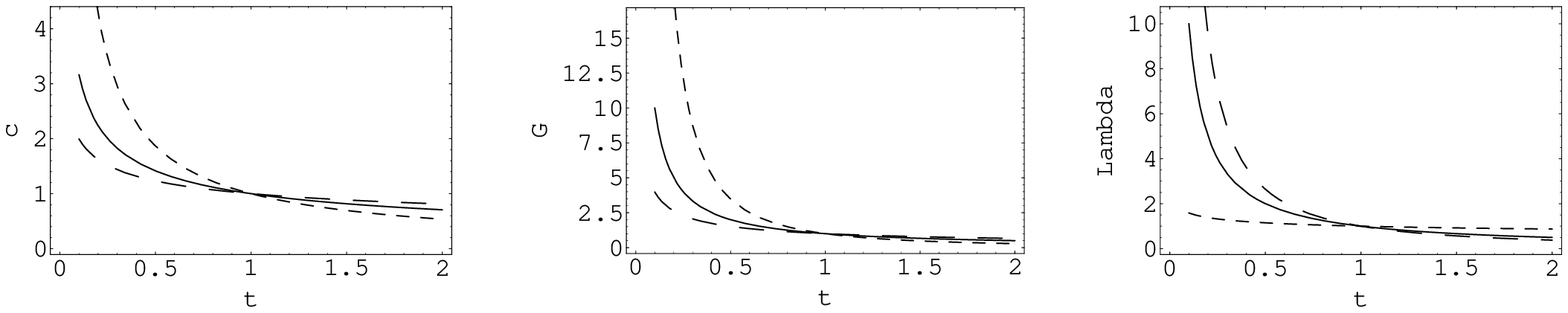}%
\caption{We see the behaviour of $G,c$ and $\Lambda$  for the first class of solutions for different values of $\alpha:\alpha=0.5$ (solid curve), $\alpha=0.9$ (dotted
curve), $\alpha=0.3$ (matter era)(dashed curve). In all cases the constants are decreasing functions.}%
\label{fig1}%
\end{center}
\end{figure}

Hence, in this case we have found that (see fig.\ref{fig1}):
\begin{equation}
G=G_{0}t^{-2\alpha},c=c_{0}t^{-\alpha},\Lambda
=\Lambda_{0}t^{-2(1-\alpha)},f=\left(  A_{\omega}t\right)
^{\frac{2}{3\left(  \omega+1\right)  }},\rho=\rho_{0}t^{-2}.%
\end{equation}
This is the solution that we have obtained with dimensional analysis in the 
previous section and we shall show this is only solution compatible with our 
previous solution$^{15}$ obtained in the framework of a bulk viscous fluid 
(full causal theory).

\subsection{\textbf{Case II}, $b=a=0$}\label{sol2}

In this case, we find that
\begin{equation}
\frac{G}{c^{2}}=\widetilde{B}t^{\varkappa},
\end{equation}
where $\varkappa=\delta-2$ and $\delta=\frac{d}{2e}.$ On following the same
procedure as above, we find that 

\begin{equation}
\frac{dt}{\xi}=\frac{d\rho}{\eta}\Longrightarrow\rho=\rho_{0}t^{-\delta},
\end{equation}
we must impose the condition $sign(d)=sign(e)$, i.e., $\delta\in\mathbb{R}^{+},$
in order that the solution has some physical meaning that the energy density is a
decreasing function of time $t$. It is observed that if $d=4e$ then we obtain
same solution that the obtained one in the case I.  The scale factor is found
to be
\begin{equation}
f=K_{f}t^{\frac{\delta}{3\left(  \omega+1\right)  }},
\end{equation}
where $K_{f}$ is an integration constant, and therefore, the Hubble parameter is:
\begin{equation}
H=\frac{\delta}{3\left(  \omega+1\right)  t},
\end{equation}
which is similar to the scale factor obtained in case I. To obtain the behaviour 
of the ``constants'' $G$, $c$ and $\Lambda$, we follow the same steps as in case I,
i.e., from%
\begin{equation}
c^{2}\Lambda=3H^{2}-\frac{8\pi G}{c^{2}}\rho,
\end{equation}
we obtain the behaviour of $\Lambda$ being:%
\begin{equation}
\Lambda=\frac{l}{c^{2}t^{2}},
\end{equation}
where, $l=\left(  K_{1}-K_{2}\right)$, $K_{1}=\frac{\delta^{2}%
}{3\left(  \omega+1\right)  ^{2}}$ and $K_{2}=8\pi\rho_{0}\widetilde{B}$ i.e., $l\in\mathbb{R}^{+}$.
Therefore,
\begin{equation}
\Lambda^{\prime}=-\frac{2l}{c^{2}t^{2}}\left(  \frac{c^{\prime}}{c}+\frac
{1}{t}\right)
\end{equation}

If we substitute all this results into the next equation
\begin{equation}
\frac{\Lambda^{\prime}c^{4}}{8\pi G\rho}+\frac{G^{\prime}}{G}-4\frac
{c^{\prime}}{c}=0,
\end{equation}
we obtain an ODE for $c$, i.e.,
\begin{equation}
\frac{c^{\prime}}{c}\left(  \lambda-2\right)  =-\left(
\lambda-2+\delta\right)\frac{1}{t}
\end{equation}
where, $\lambda=\left(  -\frac{l}{4\pi\rho_{0}\widetilde{B}}\right)$, $\lambda\in\mathbb{R}^{-}$, which
leads to
\begin{equation}
c=c_{0}t^{-\alpha}%
\end{equation}
with $\alpha=\left( 1+\frac{\delta}{\lambda-2}\right) $ scuh that $\alpha\in[0,1)$. In this way
we can find the rest of quantities:%
\begin{equation}
G=G_{0}t^{-2\left(  \alpha+1\right) +\delta },\quad{
\ \ \ }\Lambda=\Lambda_{0}t^{-2\left(  1-\alpha\right)  },
\end{equation}
note that $\alpha<1$. The case $\alpha=1\Longleftrightarrow\delta=0$ is forbiden and $\alpha=0$ brings us to the limiting case of the $G,\Lambda$ variable cosmologies$^{41}$.

\begin{figure}[h]
\begin{center}
\includegraphics[height=2.6835in,width=4.3336in]{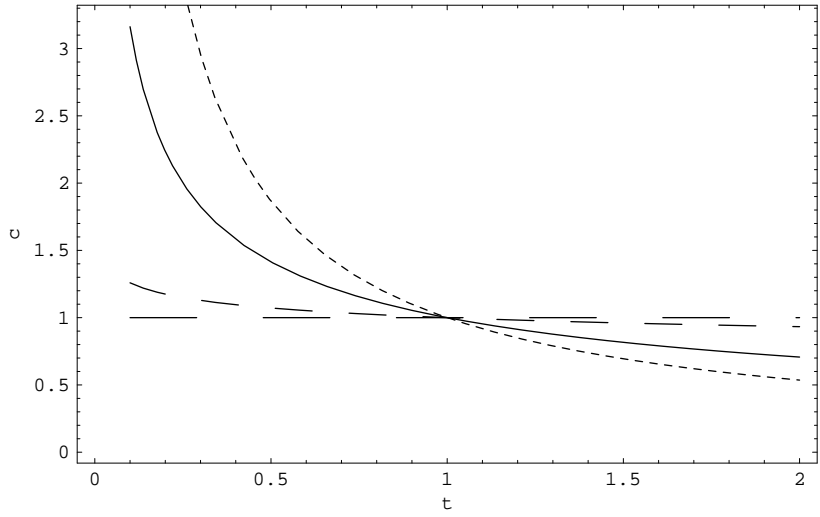}
\caption{Time variation of $c(t)$ for the second class of solutions for
different values of $\alpha:\alpha=0.5$ (solid curve), $\alpha=0.9$ (dotted
curve), $\alpha=0.1$ (dashed curve) and $\alpha=0.000001$ (long dashed curve),
the last solution describes the case $c(t)=const.$}%
\label{fig2}%
\end{center}
\end{figure}

\begin{figure}[h]
\begin{center}
\includegraphics[height=2.6852in,width=4.3318in]{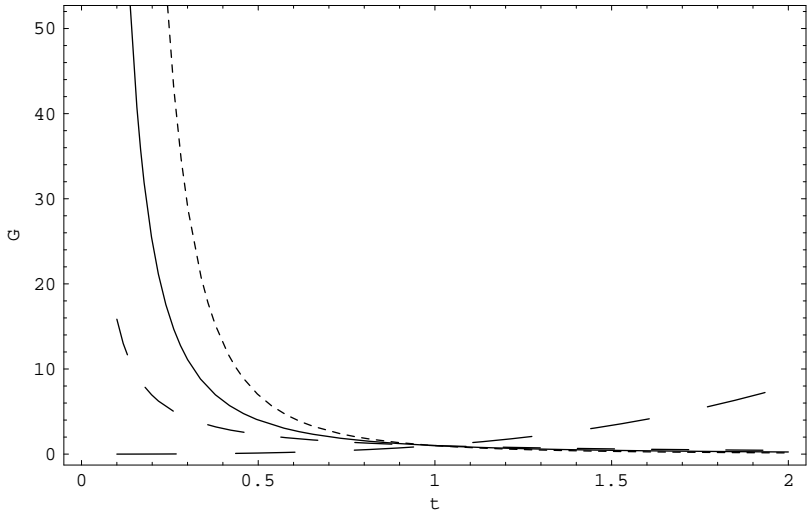}%
\caption{The variation of the gravitational ``constant'' $G(t)$, for
different values of $\alpha$ and $\delta.$ $:\alpha=0.5$ and $\delta=1$ (solid
curve), $\alpha=0.9$ and $\delta=1$ (dotted curve), $\alpha=0.1$ and
$\delta=1$ (dashed curve) and $\alpha=0.000001$ and $\delta=5$ (long dashed
curve), the last curve describes a growing solution.}%
\label{fig3}%
\end{center}
\end{figure}

\begin{figure}[h]
\begin{center}
\includegraphics[height=2.6844in,width=4.3474in]{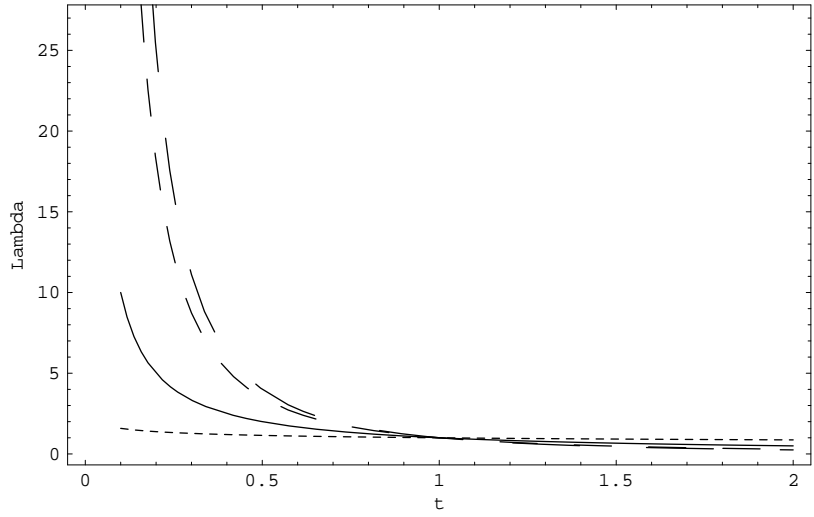}%
\caption{Time variation of $\Lambda(t)$ for the second class of solutions for
different values of $\alpha:\alpha=0.5$ (solid curve), $\alpha=0.9$ (dotted
curve), $\alpha=0.1$ (dashed curve) and $\alpha=0.000001$ (long dashed curve).
In all cases, $\Lambda(t)$ is a decreasing function.}%
\label{fig4}%
\end{center}
\end{figure}

We notice that this solution is very similar to the case I but in this case all
the parameters are perturbed by $\delta$ and more important is the result, 
$\frac{G}{c^{2}}=\widetilde{B}t^{\varkappa}$ (see figs.\ref{fig2}, \ref{fig3} and \ref{fig4}).

\subsection{\textbf{Case III}, $b=e=0$}\label{sol3}

Following the same procedure as above, we find in this case that such
restrictions imply $\xi(t,\rho)=a,$ $\eta(t,\rho)=d\rho$ and therefore:
\begin{equation}
\frac{G^{\prime}}{G}=2\frac{c^{\prime}}{c}-\frac{d}{a},
\end{equation}
which brings us to:
\begin{equation}
\frac{G}{c^{2}}=K\exp(-\alpha t),
\end{equation}
where $\frac{d}{a}=\alpha$ and note that $\left[  K\right]  = \left[  B\right]
$ i.e has the same dimensional equation,%
\begin{equation}
\frac{dt}{\xi\left(  t,\rho\right)  }=\frac{d\rho}{\eta\left(  t,\rho\right)
}\Longrightarrow\frac{dt}{a}=\frac{d\rho}{d\rho}\Longrightarrow\rho=\rho
_{0}\exp\left(  \alpha t\right)  ,
\end{equation}
this expression only has sense if $\alpha\in\mathbb{R}^{-},$ note that
$\left[  \alpha\right]  =T^{-1}.$

The scale factor $f$ satisfies the relationship:
\begin{equation}
\rho=A_{\omega}f^{-3\left(  \omega+1\right)  }\Longrightarrow f=K_{f}%
\exp\left(  \alpha t\right)  ^{\frac{-1}{3\left(  \omega+1\right)  }},
\end{equation}
that is to say, it is a growing function without singularity. In this way, we find that
\begin{equation}
H=-\frac{\alpha}{3\left(  \omega+1\right)}=conts.\quad\quad H>0.
\end{equation}
The cosmological ``constant'' is obtained as
\begin{equation}
c^{2}\Lambda=\frac{\alpha^{2}}{3\left(  \omega+1\right)  ^{2}}-8\pi K\rho
_{0}\Longrightarrow c^{2}\Lambda=l,
\end{equation}
note that $\left[  l\right]  =T^{-2},$ if we replace all these results into
eq. (\ref{field4}) then we shall obtain the exact behaviour for $c,$ i.e.
\begin{equation}
\left(  \frac{l}{8\pi K\rho_{0}}+2\right)  \frac{c^{\prime}}{c}=\alpha,
\end{equation}
and hence,
\begin{equation}
c=K\exp(c_{0}t),
\end{equation}
where $c_{0}=\frac{\alpha}{\left(  \frac{l}{8\pi K\rho_{0}}+2\right)  }$ with $c_{0}\in\mathbb{R}^{-}$ since $\alpha\in\mathbb{R}^{-}$, that is, $c$ is a decreasing function on time $t.$

In this case, we have found
\begin{eqnarray}
c &=&K\exp(c_{0}t),\\
G &=&G_{0}\exp(\left(  -\alpha+2c_{0}\right)  t),\\
\Lambda   &=&l\exp(c_{0}t)^{-2},
\end{eqnarray}
therefore the solutions for this case are (see fig. \ref{fig3}):%
\begin{eqnarray}
G   &=&G_{0}\exp(\left(  -\alpha+2c_{0}\right)  t), c=K\exp
(c_{0}t),\Lambda=l\exp(c_{0}t)^{-2},\\
\rho &=&\rho_{0}\exp\left(  \alpha t\right),f=K_{f}\exp\left(
\alpha t\right)  ^{\frac{-1}{3\left(  \omega+1\right)  }}.
\end{eqnarray}

\begin{figure}[h]
\begin{center}
\includegraphics[height=2.1871in,width=6.1445in]{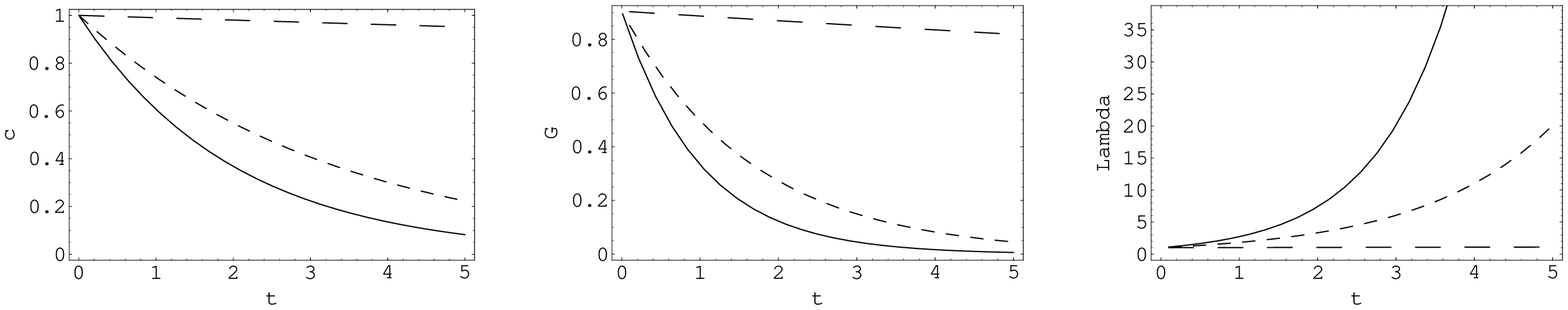}%
\caption{We see the behaviour of constants for the third class of solutions for
different values of $c_{0}:c_{0}=-0.5$ (solid curve), $c_{0}=-0.3$  (dotted
curve), $c_{0}=-0.01$  (dashed curve) and $\alpha=-0.1$.
In all cases, $\Lambda(t)$ is a growing function.}%
\label{fig5}
\end{center}
\end{figure}

As we can see, equation (\ref{ber4}) allows us to obtain more cases, 
however, we examine their physical validity by considering matter creation in 
our model in the next section. The behavior of this class of solutions for
different values of $c_0$ is shown in figure \ref{fig5}.

\section{Matter Creation}

As we have seen in the previous section, solutions \ref{sol2} and \ref{sol3} 
suggest us a new scenario since these solutions may describe the early 
universe in a very different way than our previous solution$^{15}$. 
For this reason, in this section, we study briefly the important case in 
which  adiabatic matter creation$^{42-45}$ can be taken into account, 
in order to get rid of the entropy problem. The matter creation theory is 
based on an interpretation of the matter energy-stress tensor in open 
thermodynamic systems, which leads to the modification of the adiabatic
energy conservation law and as a result including the irreversible matter
creation. The matter creation corresponds to an irreversible energy flow
from the gravitational field to the constituents of the particles created
and this involves the addition of a creation pressure $p_c$ in the matter
energy-momentum tensor which we discuss below.

The field equations that now govern our model are as follows: 
\begin{eqnarray}
2H^{\prime }+3H^{2} &=&-\frac{8\pi G(t)}{c^{2}(t)}(p+p_{c})+c^{2}(t)\Lambda
(t),  \label{m1} \\
3H^{2} &=&\frac{8\pi G(t)}{\,c^{2}(t)}\rho +c^{2}(t)\Lambda (t),  \label{m2}
\\
n^{\prime }+3nH &=&\psi ,  \label{m3}
\end{eqnarray}
and taking into account our general assumption i.e. 
\begin{equation}
T_{i;j}^{j}-\left( \frac{4c_{,j}}{c}-\frac{G_{,j}}{G}\right) T_{i}^{j}+\frac{%
c^{4}(t)\delta _{i}^{j}\Lambda _{,j}}{8\pi G}=0,
\end{equation}
with $T_{i;j}^{j}=0,$ we obtain the two equations 
\begin{equation}
\rho ^{\prime }+3\left( \rho +p+p_{c}\right) H=0,  \label{maciza1}
\end{equation}
and 
\begin{equation}
\frac{\Lambda ^{\prime }c^{4}}{8\pi G\rho }+\frac{G^{\prime }}{G}-4\frac{%
c^{\prime }}{c}=0,  \label{maciza2}
\end{equation}
where $n$ is the particle number density, $\psi $ is the function
that measures the matter creation, $H=f^{\prime }/f$ represents the Hubble
parameter ($f$ is the scale factor that appears in the metric), $p$ is the
thermostatic pressure, $\rho $ is energy density and $p_{c}$ is the pressure
that generates the matter creation.

The creation pressure $p_{c}$ depends on the function $\psi $. For adiabatic
matter creation this pressure takes the following form: 
\begin{equation}
p_{c}=-\left[ \frac{\rho +p}{3nH}\psi \right] .  \label{w2}
\end{equation}
The state equation that we next use is the well-known expression 
\begin{equation}
p=\omega \rho,  \label{w3}
\end{equation}
where $\omega =const.$  and $\omega \in (-1,1]$. We assume that the matter
creation function follows the law$^{45}$: 
\begin{equation}
\psi =3\beta nH,  \label{w5}
\end{equation}
where $\beta $ is a dimensionless constant (if $\beta =0$
then there is no matter creation since $\psi =0)$. The generalized principle
of conservation $T_{i;j}^{j}=0,$ for the stress-energy tensor (\ref{maciza1}%
) leads us to: 
\begin{equation}
\rho ^{\prime }+3(\omega +1)\left( 1-\beta \right) \rho H=0.  \label{w4}
\end{equation}

Therefore, the new set of field equations are: 
\begin{eqnarray}
2H^{\prime }+3H^{2} &=&-\frac{8\pi G(t)}{c^{2}(t)}(p+p_{c})+c^{2}(t)\Lambda
(t),  \label{ma1} \\
3H^{2} &=&\frac{8\pi G(t)}{\,c^{2}(t)}\rho +c^{2}(t)\Lambda (t),  \label{ma2}
\\
\rho ^{\prime }+3(\omega +1)\left( 1-\beta \right) \rho H &=&0,  \label{ma3}
\\
\frac{\Lambda ^{\prime }c^{4}}{8\pi G\rho }+\frac{G^{\prime }}{G}-4\frac{%
c^{\prime }}{c} &=&0.  \label{ma4}
\end{eqnarray}

Now, using the same procedure that was followed in section 4, we find the 
new equation on which we can apply the Lie method. We rewrite the field
equations as follows: from (\ref{ma1}) $-$ (\ref{ma2}) we obtain 
\begin{equation}
2\frac{f^{\prime \prime }}{f}-2\left( \frac{f^{\prime }}{f}\right) ^{2}=-%
\frac{8\pi G}{c^{2}}\left( p+p_{c}+\rho \right) ,
\end{equation}
and therefore 
\begin{equation}
2\left( H\right) ^{\prime }=-\frac{8\pi G}{c^{2}}\left( p+p_{c}+\rho \right).
\end{equation}
From (\ref{ma3}), we can obtain 
\begin{equation}
H=-\frac{\rho ^{\prime }}{3\left( \omega +1\right) (1-\beta )\rho },
\end{equation}
therefore 
\begin{equation}
\left( \frac{\rho ^{\prime }}{\rho }\right) ^{\prime }=12\pi \left( \omega
+1\right) ^{2}(1-\beta )^{2}\frac{G}{c^{2}}\rho,
\end{equation}
making $12\pi \left( \omega +1\right) ^{2}(1-\beta )^{2}=\tilde{A}$ and expanding, 
we obtain 
\begin{equation}
\rho ^{\prime \prime }=\frac{\rho ^{\prime 2}}{\rho }+\tilde{A}\frac{G}{c^{2}%
}\rho ^{2}.  \label{newlie}
\end{equation}

Therefore, we obtain the same equation that we obtained in section $4$, namely
equation (\ref{ber1}) except the constant $\tilde{A}$. Hence, within this framework
the entropy and horizon problems are solved for our model.

\section{Conclusions.}

In this paper we have studied the behaviours of time-varying ``constants'' 
$G,c$ and $\Lambda$ in a perfect fluid model. We began reviewing the scaling 
solution obtained through dimensional analysis. He have shown that this 
solution connects with our previous solution$^{15}$ where we studied the 
behaviour of the ``constants'' $G,c$ and $\Lambda$ in a full causal bulk 
viscous model arriving to the conclusion that this model tends to a perfect 
fluid one when we impose the condition $\omega=0$ in the equation of state
(dust solution). 

To obtain this solution, we imposed the assumption, $div(T_{j}^{i})=0$, from 
which we obtained the dimensional constant $A_{\omega}$ that relates 
$\rho\propto f^{-3(\omega+1)}$ and the relationship $G/c^{2}=const.=B$ 
remaining constants for all value of $t,$ i.e. $G$ and $c$ vary but in 
such a way that $G/c^{2}$ remain constant. With these two hypothesis, we 
have obtained the scaling solution that connects perfectly with one obtained 
in our previous paper$^{15}$, i.e. with the bulk viscous model in the 
matter-dominated era.

In this context, the solution obtained through dimensional analysis show us
that the ``constants'' $G,c$ and $\Lambda$ are decreasing functions of time,
but in this case decrease slowly than in the radiation predominance era, while
$\rho$ and $f$ behave as in the FRW model solvingthe horizon problem.

Since we have been able to found a solution through similarity, i.e. through
dimensional analysis, it is possible that there are other symmetries of the
model, since dimensional analysis is a reminiscent of scaling symmetries,
which obviously are not the most general form of symmetries. Therefore, we
studied the model through the method of Lie group symmetries, showing that
under the assumed hypotheses, there are other solutions of the field equations.

The first solution obtained is the already obtained one through similarity, but,
in this case we have showed the condition $G/c^{2}$ arises as a result and not as 
an {\it ad-hoc} condition. We also have studied two other cases which can be 
considered as physically relevant solutions since $f$ is a growing
function on time and $\rho$ is a decreasing function on time. They could 
describe very early cosmological solutions (inflationary ones) but in this 
context we cannot solve the entropy problem. We have considered matter creation
and and using a recasted set of equations, shown that when matter creation is 
taken into account, the horizon and entropy problems are solved for the two 
solutions.

\end{document}